\def\pdot {\dot P}
\def\ltsima{$\; \buildrel < \over \sim \;$}
\def\lsim{\lower.5ex\hbox{\ltsima}}
\def\gtsima{$\; \buildrel > \over \sim \;$}
\def\gsim{\lower.5ex\hbox{\gtsima}}
\def\msole{~M_{\odot}}
\def\mdot {\dot M}
\def\uu {4U~0142$+$61~}
\def\oo {1E~1048.1$-$5937~}
\def\kes {1E~1841$-$045~}
\def\axj {AX~J1845.0$-$0300~}
\def\rx {1RXS~J170849$-$400910~}
\def\ee {1E~2259$+$586~}

\documentstyle[editedvolume,psfig]{crckapb} 
 
\begin{opening}
\title{THE ANOMALOUS X-RAY PULSARS}

\author{SANDRO MEREGHETTI}
\institute{Istituto di Fisica Cosmica G.Occhialini - CNR\\
           via Bassini 15, I-20133 Milano, Italy}

\end{opening}

\runningtitle{THE ANOMALOUS X-RAY PULSARS}

\begin{document}

{\it
Invited Review presented at the NATO Advanced Study Institute\\
{\bf ``The Neutron Star - Black Hole Connection''}\\
Elounda, Crete - 7-18 June 1999.
}



\section{Introduction}

In the last few years it has been recognized that a few X--ray
pulsars, which are not rotation powered, have peculiar properties that
sets them apart from the majority of accreting pulsars in X--ray binaries.
These objects, initially suggested as a homogeneous new class of pulsators
in 1995 (Mereghetti \& Stella 1995), have been named in different ways,
reflecting our ignorance on their true nature: Very Low Mass X--ray 
Pulsars, Braking Pulsars, 6-sec  Pulsars, Anomalous X--ray Pulsars. 
The latter designation (AXP) has become the most popular and will be used
here. 

Though we can be reasonably confident that the AXP are rotating 
neutron stars without massive companions,  
it is unclear whether they are solitary objects or are
in binary systems with very low mass stars. 
As a consequence, different mechanisms for powering their X--ray 
emission have been proposed, involving either accretion or other less
standard processes such as, e.g.,  the decay of magnetic energy.

\begin{table}[]
\begin{tabular}{|c|c|c|c|c|}
 
\hline

{\bf  SOURCE}       &{\bf P (s)}  &{\bf $\pdot$  (s s$^{-1}$)}  &{\bf SNR}       &  {\bf    SPECTRUM  }  \\
                   &             &                    & {\bf d (kpc)/age (kyr)}  &  {\bf kT$_{BB}$/$\alpha_{ph}$}    \\
\hline
\multicolumn{5}{|c|}{ {\bf Anomalous X--ray Pulsars (AXP)} } \\
\hline
1E~1048.1--5937    & 6.45 &[1.5--4]$\times$10$^{-11}$ &    --                 & BB+PL  [3] \\
                   & [1]  &      [2,3]                &                       & $\sim$0.64 keV / $\sim$2.5\\
1E~2259+586        & 6.98 & $\sim$5$\times$10$^{-13}$ & G109.1--0.1 [7,8,9]   & BB+PL [9]                    \\
                   & [4]  &       [5,6]               &   4--5.6 / 3--20      & $\sim$0.44 keV / $\sim$3.9\\
4U~0142+61         & 8.69 & $\sim$2$\times$10$^{-12}$ &    --                 & BB+PL [11,12]                \\
                   & [10] &       [11]                &                       & $\sim$0.4 keV /  $\sim$4   \\
RXSJ170849--4009   &11.00 & 2$\times$10$^{-11}$       &    --                 & BB+PL [13]                \\
                   & [13] &       [14]                &                       & $\sim$0.41 keV/ 2.92  \\
1E~1841--045       &11.77 & 4.1$\times$10$^{-11}$     &  Kes 73 [17,18]       & PL  [19]                \\
                   & [15] &     [16]                  &   6--7.5 / $\lsim$3   &  -- / $\sim$3.4  \\
AX~J1845.0--0300   & 6.97 & --                        & G29.6+0.1 [21]        & BB  [20]    \\
                   & [20] &                           &     $<$20 / $<$8      & $\sim$0.7 keV / --  \\

\hline
\multicolumn{5}{|c|}{{\bf Pulsed Soft Gamma-ray Repeaters (SGR)}} \\
\hline
SGR~0526--66   &  8.1    &  --                             &  N49 in LMC [23]     &   uncertain [24]   \\
               &  [22]   &                                 &                      &               \\
SGR~1806--20   &  7.48   & $\sim$8.3$\times$10$^{-11}$     & G10.0--0.3 [26]      & PL  [27]    \\
               &  [25]   &        [25]                     &                      & $\sim$2.2   \\
SGR~1900+14    &  5.16   & $\sim$[5--14]$\times$10$^{-11}$ & G42.8+0.6 [31]       &    BB+PL [32]    \\
               &  [28]   &       [29,30]                   &                      & $\sim$0.5 keV / 1.1 \\
 \hline
\multicolumn{5}{|c|}{{\bf (Candidate) Radio-Quiet Neutron Stars  } } \\
\hline
1E~1207--5209 [33]    &   --    &  --                          & G296.5+10      &    BB [33]           \\
                      &         &                              &                &   $\sim$0.25 keV \\
1E~1614--5055 [34]    &   --    &  --                          & RCW 103        &    BB [35]         \\
                      &         &                              &                &   $\sim$0.6 keV \\
1E~0820--4247 [36]    & 0.075 ? & 1.5 10$^{-13}$            ?  & Puppis A       &    BB [35] \\
                      &  [37]   &         [37]                 &                &   $\sim$0.3 keV  \\
RX~J0720.4--3125      &  8.39   & --                           &   --           &    BB [38]             \\
                      &  [38]   &                              &                &   $\sim$0.08 keV \\
RXJ1856.5--3754 [39]  &   --    & --                           &        --      &    BB [39]        \\
                      &         &                              &                &   $\sim$0.06 keV \\
\hline
\end{tabular}

[1] Seward et al. 1986;
[2] Mereghetti 1995;
[3] Oosterbroek et al. 1998; 
[4] Fahlman \& Gregory 1981;
[5] Baykal \& Swank 1996;
[6] Kaspi et al. 1999;
[7] Hughes et al. 1984;
[8] Rho \& Petre 1997;
[9] Parmar et al. 1998;
[10] Israel et al. 1994;
[11] Israel et al. 1999a;
[12] White et al. 1996;
[13] Sugizaki et al. 1997;
[14] Israel et al. 1999b;
[15] Vasisht \& Gotthelf 1997;
[16] Gotthelf et al. 1999;
[17] Sanbonmatsu \& Helfand 1992;
[18] Helfand et al. 1994;
[19] Gotthelf \& Vasisht 1997;
[20] Torii et al. 1998;
[21] Gaensler et al. 1999;
[22] Mazets et al. 1979;
[23] Cline et al. 1982;
[24] Marsden et al. 1996;
[25] Kouveliotou et al. 1998;
[26] Kulkarni et al. 1994;
[27] Sonobe et al. 1994;
[28] Hurley et al. 1999;
[29] Kouveliotou et al. 1999;
[30] Woods et al. 1999b;
[31] Vasisht et al. 1994;
[32] Woods et al. 1999a;
[33] Mereghetti et al. 1996;
[34] Tuohy \& Garmire 1980;
[35] Gotthelf et al. 1997;
[36] Petre et al. 1996;
[37] Pavlov et al. 1999;
[38] Haberl et al. 1997;
[39] Walter et al. 1996;
\centerline{\bf~~}
\centerline{\bf Table 1 - AXP and related objects}

\end{table}

The properties that 
distinguish the AXP from the more common   pulsars  found in 
High Mass X--Ray Binaries (HMXRB) are the following:

a)  spin periods in a narrow range ($\sim$6-12 s),
compared to the much broader one (0.069 - $\sim$10$^4$ s)
observed   in HMXRB pulsars (see Fig.~1)
 
b) no identified optical counterparts, with upper limits excluding 
the presence of normal massive companions, like OB (super)giants and/or Be stars

c) very soft X--ray spectra (characteristic temperature $\lsim$ 1 keV and/or
power-law photon index $\gsim$ 3)
 
d) relatively low   X--ray luminosity  
($\sim10^{34}$-10$^{36}$ erg s$^{-1}$) compared to that of HMXRB pulsars
(see Fig.~1)

e) little or no  variability (on timescales from hours to years)

f)  relatively stable spin period evolution,
with long term spin-down trend

g) a few of them are associated with supernova remnants.

There are now six members of the AXP class (section 2).
This review is mainly focussed on their observational properties (section 3),
while the models are     briefly discussed 
in section 4.

\begin{figure}
\centerline{\psfig{figure=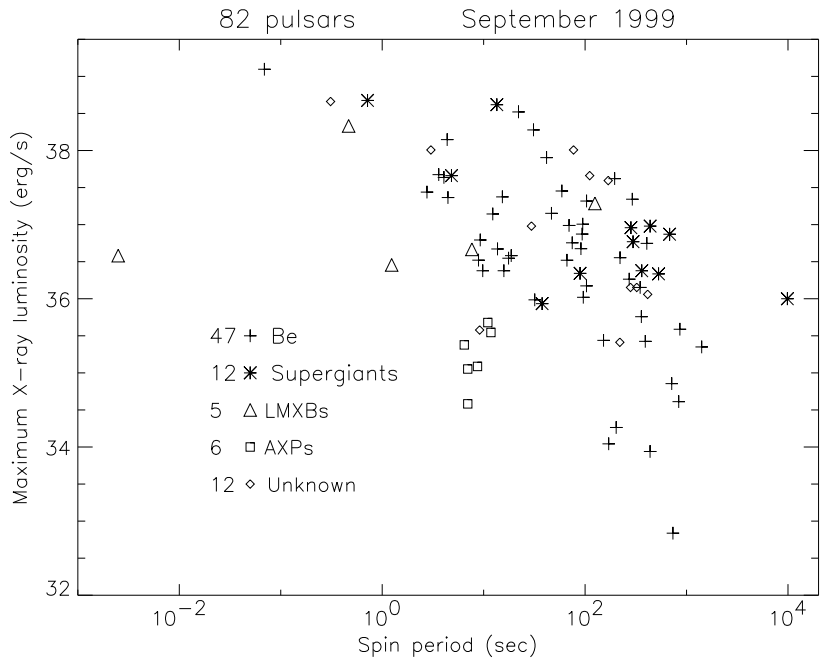 ,width=12cm,height=9cm,angle=90} }
\caption{Spin period and maximum X--ray luminosity of different classes of X--ray pulsars.
(adapted from Tiengo 1999).}
\end{figure}

\section{The AXP sample}

Table 1 lists the 6 pulsars that share the above characteristics
and form the current AXP sample. For comparison, also the properties
of other objects that might be related to the AXP are reported in 
Table 1. The soft gamma-ray repeaters (SGR) have P and $\pdot$ values
very similar to those of the AXP. As discussed below, the magnetar model,
originally developed to explain the SGR, has also been applied to the
AXP. A few other (candidate) isolated neutron stars have some similarities
with the AXP  (see Fig.~2), but more observations are needed to establish their
nature. 

On the basis of a better understanding of the AXP properties and/or
of new observational results, we
exclude from the AXP group a  few sources  that have been previously
considered as part of this class of objects.
4U~1626--67 was originally included in the  AXP class (Mereghetti \& Stella 1995),
but several authors pointed out its different nature: 
it has a   harder spectrum,  an optical identification,
there is clear evidence for a binary nature,
and showed an extended period of spin-up 
(van Paradijs et al. 1995,
Ghosh et al. 1997). 

The presence of pulsations at   5.45 s in the ROSAT source
RXJ~1838.4--0301 (Schwentker 1994)
has not been confirmed by more sensitive ASCA observations. 
Furthermore, optical observations of its possible 
counterparts revealed the presence
of a  main sequence K5  star with V$\sim14.5$  (Mereghetti, Belloni \& Nasuti 1997).
This star could be responsible for the observed X--ray flux, 
since the implied X--ray to optical flux ratio (f$_x$/f$_{opt}$)
is compatible with the level of coronal emission expected  
in late type stars. Thus it seems very likely that  RXJ~1838.4--0301
is not a pulsar -- i.e. the statistical significance of the periodicity
was overestimated  (Schwentker 1994).   

The 8.4 s pulsar RX~J0720.4--3125 (Haberl et al. 1997) has also been sometimes
included in the AXP group, on the basis of its period value, high f$_x$/f$_{opt}$,
and soft spectrum. Indeed its spectrum is
even softer than that of AXP and it can only be detected thanks to the 
very low interstellar absorption (N$_{H}\sim$10$^{20}$~cm$^{-2}$). 
Since this is taken as evidence
for a very small distance (d$\sim$100~pc), the implied luminosity 
of $\sim$3$\times$10$^{31}$ erg s$^{-1}$ is
much smaller than that of the AXP. It has been suggested that  RX~J0720.4--3125
is an old neutron star accreting from the interstellar medium, 
but the possibility of a medium age neutron star, still emitting through
dissipation of its internal heat, cannot be excluded.

\section{Observational Properties of the AXP}

\subsection{Spectra }

The AXP are characterized by soft X-ray spectra, clearly different from those 
of the pulsars in HMXRB. The latter have relatively hard spectra in the 2-10 
keV range (i.e. power law photon index $\alpha_{ph}\sim$1) that steepen with an 
exponential cut-off above $\sim$20 keV. On the contrary, since their first 
observations, AXP showed very soft power law spectra, 
with  $\alpha_{ph}\gsim$3-4.
Reports of possible cyclotron features at low energy ($\sim$5--10 keV) 
in \ee (Iwasawa et al. 1992)  have not been confirmed.
 
Recent observations with ASCA and BeppoSAX, have shown that in most cases a single 
power law is not sufficient to describe the spectra of AXP. All the AXP for 
which good quality observation are available (White et al. 1996,
Parmar et al. 1998, Oosterbroek et al.1998, 
Israel et al. 1999a) require the combination of a blackbody-like component 
with kT$\sim$0.5 and a steep power law ($\alpha_{ph}\sim$3--4).
A single power law is adequate to describe the spectrum of \kes
(Gotthelf \& Vasisht 1998), but the analysis is complicated by the
presence of the underlying emission from  the SNR that might hamper the
detection of  the blackbody component.
In \axj a   blackbody with kT$\sim$0.7 keV gives a good fit
without the need for an additional power law component (Gotthelf \& Vasisht 1997).

The spectral parameters for all the AXP are summarized in 
Table 2.  The emitting area inferred from the blackbody components, that 
account up to  $\sim$40-50\% of the observed luminosity, is compatible with a 
large fraction of a neutron star surface. 

Some evidence for spectral variations as a function 
of the spin-period phase has been reported for
several AXP:  \ee (Iwasawa et al. 1992,
Corbet et al. 1995, Parmar et al. 1988), \uu (Israel et al. 1999a), 
\oo (Corbet \& Mihara 1997, Oosterbroek et al. 1988) and \rx (Sugizaki et al. 1997).
Unfortunately, the relatively poor energy resolution, and
the limited statistics, do not allow to unambiguously
characterize the spectral variations in the two 
separate components.

It is possible that this two component model   
be an oversimplified description of the true underlying spectra resulting 
from the current instrumental limitations. 
Future observations with XMM should resolve this issue, possibly leading to the 
discovery of narrow spectral features that so far escaped detection. Note in 
particular that the energy of cyclotron lines from ions lies in the 0.1 - 10 keV 
range for the high values of the magnetic field (B$\sim$10$^{14}$) expected 
for the magnetar model (see section 4.2). 

\begin{figure}
\centerline{\psfig{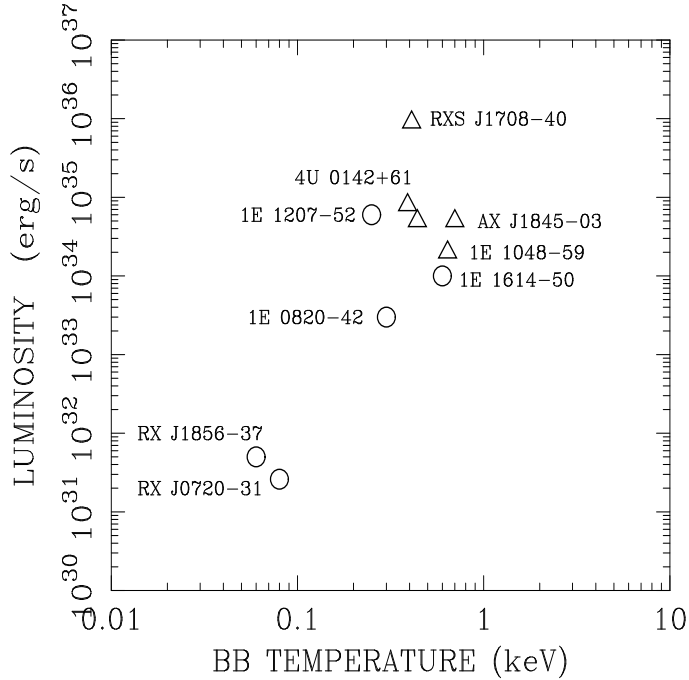} }
\caption{
''Hertzsprung-Russell diagram`` for AXP (triangles) and isolated neutron stars (circles).
The temperatures refer to the blackbody spectral components. }
\end{figure}

\subsection{ Distances and luminosities }

Due to the lack of optical identifications, the distances of AXP are 
quite uncertain (with the exception of the two in SNR, section 3.8).
However, some   constraints can be derived from 
their location in the Galaxy.
 
The low distribution on the galactic plane  ($<|$b$|>$=0.35$^{\circ}$), indicates
that, as a population, they are unlikely to be nearby ($\lsim$ 1 kpc) objects. 
Such a conclusion is also consistent with the relatively high
column density derived from the  X--ray spectral fits (Table 2).

\oo lies in the direction of the Carina Nebula, which is thought
to contribute to the high absorption measured in its spectrum,
giving a lower limit to the distance of 2.8 kpc (Seward et al. 1986). 
A similar argument can be made for \uu that probably lies behind
a local (d$\lsim$1 kpc) molecular cloud clearly visible in absorption on the  
Palomar Sky Survey plate
(Israel, Mereghetti \& Stella 1994).
On the other hand, a distance much in excess of $\sim$5 kpc, would
place this source outside the Galaxy.

The two AXP associated with SNR  have better distance estimates:
6-7.5 kpc for \kes in Kes 73  (Sanbonmatsu \& Helfand 1992)
and  5.6 kpc for \ee in G109.1--0.1 (Hughes et al. 1984).

\rx is in the
general direction of the galactic center region and has a 
highly absorbed X--ray spectrum, which   suggests
a distance of the order of 8 kpc or more. 

According to Torii et al. (1998), \axj could be located in the Scutum arm,
at d$\sim$8.5 kpc. Also this source is very absorbed and its
distance could be larger. 
More information will be obtained if its association
with the new radio SNR
found by Gaensler et al. (1999)  is confirmed.

Based on these distances and the observed fluxes,   luminosities 
in the $\sim10^{34}$-10$^{36}$ erg s$^{-1}$  range are  obtained for the AXP
(see Table 2).  

Another uncertainty affecting the AXP luminosity estimates is the
correction for the (model dependent) X--ray absorption. 
In principle, this could be a relevant factor, due to the steepness
of the observed spectra. Note in fact that for a  power law spectrum
that extends down to low energy with, e.g., $\alpha_{ph}\sim$4,
the flux in the 0.5-2 keV range is $\sim$15 times the 2-10 keV one.
However, for the blackbody plus power law spectra discussed in section 3.1
this correction is much smaller.
It seems therefore well established that AXP have
X--ray luminosities smaller than those typically observed in persistent
HMXRB pulsars.

\begin{table}[htb]
\begin{tabular}{|c|c|c|c|c|}
 
\hline
 SOURCE            & L$_{x}^{~~(a)}$  & kT$_{BB}^{(c)}$    & $\alpha_{ph}^{(e)}$   &  L$_{BB}$/L$_{tot}$      \\
                   &  d$^{(b)}$       &  R$_{BB}^{(d)}$    &     N$_H$             &  Pulsed Fraction           \\
\hline
1E~1048.1--5937    & 2 10$^{34} erg s^{-1}$  &  0.64 $keV$ &          2.5           &  0.55        \\
                   &          5 $kpc$        &  1    $km$  &   5 10$^{21} cm^{-2}$  &  $\sim$70\%  \\
1E~2259+586        & 5 10$^{34} erg s^{-1}$  &  0.44 $keV$ &          3.9           &  0.4         \\
                   &          5 $kpc$        &  4.1  $km$  &   9 10$^{21} cm^{-2}$  &  $\sim$30\%  \\
4U~0142+61         & 8 10$^{34} erg s^{-1}$  &  0.4  $keV$ &           4            &  0.4         \\
                   &          1 $kpc$        &  2.4  $km$  & 1.1 10$^{22} cm^{-2}$  &  $\sim$10\%  \\
1RXSJ170849--4009  & 9 10$^{35} erg s^{-1}$  &  0.41 $keV$ &          2.92          &  0.17        \\
                   &          8 $kpc$        &  3.2  $km$  & 1.4 10$^{22} cm^{-2}$  &  $\sim$30\%  \\
1E~1841--045       & 3 10$^{35} erg s^{-1}$  &   --        &         3.4            &   --         \\
                   &          7 $kpc$        &             &   3 10$^{22} cm^{-2}$  &  $\sim$35\%  \\
AX~J1845.0--0300   & 5 10$^{34} erg s^{-1}$  &  0.7  $keV$ &          --            &   --         \\
                   &          8 $kpc$        &  1.5  $km$  & 4.6 10$^{22} cm^{-2}$  &  $\sim$50\%  \\

\hline
\end{tabular}

(a) corrected for interstellar absorption \\
(b) assumed values, see section 3.2 for the uncertainties \\
(c) temperature of blackbody component\\
(d) equivalent radius of blackbody component \\
(e) photon index of power law component

\centerline{\bf~~}
\centerline{\bf Table 2 - Spectral properties of AXP}

\end{table}

\subsection{Variability}

In general, AXP have  relatively steady X--ray fluxes, compared with the kind
of variability displayed by other classes of accreting compact objects. 
Most   AXP have been detected at similar
flux levels by all the satellites that looked at them.
There are, however, some interesting exceptions. 

The best evidence for flux variability has been so far obtained for \axj. 
This source was discovered  at a flux level 
of 4.2$\times$10$^{-12}$ erg cm$^{-2}$ s$^{-1}$ (2-10 keV)  in an  ASCA pointing performed in December
1993, but it was not visible  
3.5 years later, implying  a flux decrease greater than a factor 14
(Torii et al. 1998).  
A further ASCA observation revealed only a weak source 
at a position consistent with 
that of the AXP (Gaensler et al. 1999). 
Though a search for pulsations could not 
be performed, due to the small number of counts, 
it is likely that this source is \axj in a low 
state, a factor $\sim$10 fainter than the 1993 level. 

In a GINGA observation   performed in 1990   (Iwasawa et al. 1992),
\ee was  a factor $\sim$2  brighter than   in previous  measurements with the
same instrument. 
During the higher intensity state  
a change in the double-peaked pulse profile (a larger difference in
the relative intensity of the two pulses) was also observed,
as well as a variation in the spin-down rate. 
Most of the  other observations of
\ee, obtained with different satellites,  yielded flux measurements
of $\sim$2-3 $\times$ 10$^{-11}$   erg cm$^{-2}$ s$^{-1}$,
consistent with the lower intensity state
(see Corbet et al. 1995, Parmar et al. 1998 and references therein).

The flux  measurements available for \oo have been summarized 
by Oosterbroek et al. (1998). They show long term variations within a factor 
$\sim$5  (possibly more if a rather uncertain upper limit obtained
with the Einstein Observatory is also considered, Seward et al. 1986).
However, the comparison of these flux measurements is  affected
by the     uncertainties deriving from the use
of different instruments.

No evidence for  significant   variability  has been reported
for the three remaining AXP: \uu, \kes and \rx.   
However, since most of the relevant observations have been obtained
with different instruments (sometimes also in different energy ranges) the 
limits that one can infer on the absence of variabilty are subject to considerable 
uncertainties.   Several   measurements
were obtained with non-imaging instruments, and the   fluxes
must be corrected for the (poorly known)
contribution from other components in the field of view 
(e.g. SNRs, diffuse galactic ridge emission, other sources, etc..),
which introduce further uncertainties.


The level of variability in AXP is of
interest since  it is expected that some emission processes 
(e.g. thermal emission from the neutron star surface), produce less variability 
than other models (e.g. those involving mass accretion, which is in general
subject to intensity fluctuations). 
More detailed searches for correlations between luminosity changes and 
spin-down variations   can support accretion models, in which fluctuations 
in the mass accretion rate  produce different torques on the rotating neutron 
star. 
Finally, the possible existence of  many transient AXP with low quiescent 
luminosities, similar to \axj, has important implications for  
the total number of AXP in the Galaxy and   their inferred birthrate.

\subsection{Spin period distribution}

As shown in Fig.~1, X--ray pulsars in massive binaries have spin periods spanning several 
orders of magnitude, from 69 ms (A~0538--67) to about 3 hr (2S~0114+65).
The concentration of periods in the narrow $\sim$6-12 s interval was
one of the properties that led to the identification of the AXP as a possibly
distinct class of objects.
It is clear, however, that  a period in this range
is not enough to qualify  a pulsar as an AXP 
(in fact there are several HMXRB with 
periods similar to those of the AXP).
If we define the  AXP   as ``pulsars with a very soft spectrum,
that are neither HMXRB nor rotationally powered neutron stars, 
and have luminosity $\sim10^{34}$-10$^{36}$ erg s$^{-1}$'', 
it turns out remarkably that  all the known objects satisfying this definition
have periods of a few seconds  and  a secular spin-down (when measured). 

Why no AXP are seen with much longer, or much shorter, periods?
There are no obvious selection effects explaining this 
narrow period distribution.
Though  a chance result due to the statistics of small 
numbers cannot be ruled out, this could be a real effect related to the 
particular characteristics and evolution of these objects. 
If this period clustering reflects the fact that the AXP are (close to) 
equilibrium rotators, one has to invoke similar magnetic fields and 
accretion rates in all the AXP.

\subsection{Period evolution }

One of the distinctive peculiarities of AXP is their long term period evolution.
In general, accreting neutron stars are expected to spin-up, due to the 
angular momentum transferred from the accreting material, often forming an 
accretion disk (see, e.g. Henrichs 1983). Indeed this is observed in many 
HMXRB pulsars in which there is evidence for an accretion disk. 
Other pulsars show alternating episodes of spin-up and spin-down, 
the origin of which is not completely understood. 
On the contrary, the spin periods of AXP are increasing at a nearly constant rate
(on timescales ranging from $\sim$2,000 to $\sim$4$\times$10$^5$ yrs). 
This behaviour has    now been observed in a few AXP for an extended period, spanning
more than two decades. 

It can immediately be seen that for these values of P and $\pdot$,
and assuming the canonical value for the momentum of inertia of a 
neutron star I=10$^{45}$ g cm$^2$, the rotational energy
loss is orders of magnitude too  small to power  the observed luminosity of AXP.

Accurate timing measurements have shown that the spin-down of AXP is not constant,
but is subject to small fluctuations (see, e.g., Iwasawa et al. 1992, Mereghetti 1995).   
Baykal \& Swank (1996) showed that the level of $\pdot$ fluctuations in 
\ee, the AXP with the largest number of period measurments,
is similar to that typically observed in neutron stars 
accreting in X--ray binaries, which is several
orders of magnitude greater than that of radio pulsars.
More recently, Kaspi et al. (1999)  have been able to
obtain a phase-coherent timing solution for RXTE observations
of \ee spanning 2.6 years. These data show
a very low level of timing noise, contrary to the
previous results that were based on sparse (not phase-coherent)
observations spanning $\sim$20 years.
Also \rx , monitored with RXTE for 1.4 yrs, was found to have a similar  
level of timing noise
(Kaspi et al. 1999), while an even more stable rotator is \kes
(Gotthelf et al. 1999).
It seems therefore that, at least on timescales of a few years,
some AXP can be very stable rotators, with a timing noise
similar to that of radio pulsars -- a finding that supports
the magnetar interpretation (see section 4.2).

\subsection{Searches for orbital periods}

No periodic intensity variations, like eclipses or dips, that might indicate the presence of 
a binary system, have been detected in  AXP. Another clear 
signature of binarity, that has been of extreme importance in the study of 
HMXRB pulsars, is the presence of orbital Doppler shifts in the  pulse 
frequency. 
The most sensitive searches for orbital Doppler shifts in AXP have been 
carried out with the RXTE satellite.   Searches for orbital periods between a few 
minutes and one day gave negative results, yielding upper limits on the 
projected  semi-major axis a$_x$sin{\it i} of $\sim$30 and $\sim$60 
light-ms for \ee and \oo respectively (Mereghetti, Israel \& Stella 
1998). Similar results were obtained by Wilson et al. (1998) for \uu.

\begin{figure}
\centerline{\psfig{figure=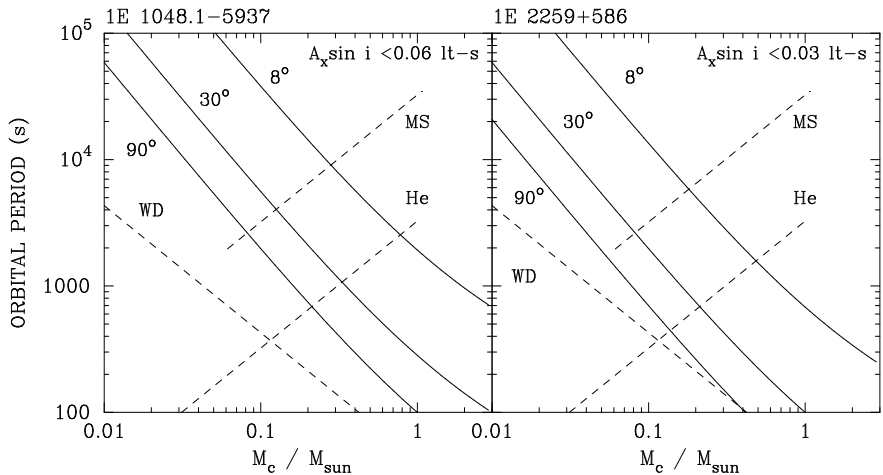 ,width=5cm,height=6cm,angle=0} }
\caption{Orbital constraints from the  $a_x\sin i$ limit   for  \oo and   \ee
(Mereghetti, Israel \& Stella 1998).
The limits on orbital period, $P_{orb}$,   
versus mass of the companion, $M_c$,  are plotted 
assuming three different values for the unknown inclination angle.
The dashed lines indicate the positions of Roche-lobe filling companions
under the assumption of conservative mass transfer driven by angular 
momentum losses due only to gravitational radiation.
They refer  to the cases of a  main sequence,  a 
He burning star and a fully degenerate hydrogen white dwarf.
Values of $P_{orb}$ and $M_c$ below the corresponding dashed line are excluded,
while those above the lines require accretion through stellar wind.
}
\end{figure}

For any   
assumed value of the inclination angle {\it i}, these limit constrain the possible 
values of the companion mass M$_c$ and orbital period (Fig.~3). As discussed by   
Mereghetti, Israel \& Stella (1998), except for the unlikely possibility that all 
these system are seen nearly face-on, main sequence companion stars can be 
ruled out. Helium burning stars with mass  M$\lsim$ 0.8 $\msole$ cannot be 
excluded, but the accretion rate produced by Roche lobe overflow would 
give a luminosity much greater than observed. 
A possibility is that of a He-burning companion, underfilling the Roche lobe 
and providing  a low rate of accretion 
through a stellar wind, as suggested by Angelini et al. (1995) for 
4U~1626-67. 
Another possibility that cannot be ruled out by the current limits 
on a$_x$sin{\it i},  and on the inferred mass accretion rates, is that of a white 
dwarf companion. For example, a white dwarf with    M$\sim$0.02$\msole$, 
filling its Roche lobe for an orbital period  of the order of $\sim$30 min, would 
give an $\mdot$ of a few $\times$10$^{-11}$ $\msole$ year$^{-1}$, consistent 
with the observed luminosities. 
Mainly due to the lack of suitable observations, no similar searches for orbital 
Doppler shifts have been performed for the three remaining AXP: \rx, \axj and \kes.

\subsection{Optical and Infrared counterparts}

The error box of \oo has a radius of 15$''$ and contains several
stars (Mereghetti, Caraveo \& Bignami 1992). Spectroscopy of the 3 brightest
objects  (V$\gsim$19)  did not yield   a plausible counterpart
showing the classical emission lines considered a signature of
accreting objects. More objects were studied by Corbet \& Mihara (1997), 
again with negative results. These studies are complicated by the
presence of diffuse H$_{\alpha}$ emission from the Carina nebula,
which affects the sky subtraction from the stellar spectra.

\ee is the AXP for which more extensive searches for counterparts
have been carried out 
(Davies \& Coe 1991, Coe \& Jones 1992, Coe et al. 1994), 
sometime leading to 
possible identifications later disclaimed by better observations.
The latest error box (5$''$ radius), reported by Coe \& Pightling (1998) contains
only three faint objects with K-band magnitudes of $\sim$18 and V$>$24.

A different situation is found for \uu, since in this case no
objects are present within the small error box ($\sim$3$''$ radius). The current best limits
are V$>$24 (Steinle et al. 1987) and  K$>$17 (Coe \& Pightling 1998).

Optical observations of the field of \rx have been reported by Israel 
et al. (1999b). These authors found  that the possible counterparts cannot
be massive early type stars, being too faint and blue 
(very distant and/or absorbed OB stars should appear more reddened by the
interstellar dust absorption).
No detailed reports on 
optical/IR  observations of \kes and
\axj have been published so far.

Though in general the limits on the possible optical/IR
counterparts of AXP allow to rule out the presence of massive companion
stars, more work is needed to explore different possibilities,
especially because it is not clear which kind of properties 
one should expect from the AXP counterparts.
Due to the crowding of these low galactic latitude fields, more precise
localizations are also needed.

\subsection{Association with supernova remnants (SNR) }

The fact that two (possibly three) AXP are found at the center of 
SNR is very important, since it gives informations on their origin,  age and 
distance. 

\ee is located close to the geometrical  center of  G109.1--0.1 (also 
known as CTB 109), a partial radio/X-ray shell with an angular diameter of 
$\sim$30$'$ (see, e.g., Rho \& Petre 1997). 
As discussed in Parmar et al. (1998), 
the estimated age for this SNR  is subject to a considerable uncertainty,
ranging from $\sim$3,000 yrs to 20,000 yrs. 
The other AXP clearly associated to a SNR is \kes. It was discovered as an 
unresolved source at the center of  Kes 73,  a young  ($\sim$ 2000 yr) SNR at a 
distance of $\sim$7 kpc (Helfand et al. 1994).  
Gaensler et al. (1999) have recently reported the discovery of a radio SNR 
around \axj .
These three AXP are found   close to the geometrical center of the respective 
SNR, implying relatively small transverse velocities for these objects.

One should not forget that three AXP  (\uu, \oo and \rx) lack visible 
SNRs.  This might indicate that the lifetime of AXP is much longer than several 
10$^4$ years. 

There are also a few unresolved X-ray sources within SNRs that, apart for the 
lack of pulsations, share the same properties of the AXP (see Table 1).
The sources in RCW 103 (Gotthelf, Petre  \& Hwank 1997),   
G296.5+10.0 (Mereghetti et al. 1996), and   Puppis A 
(Petre et al. 1996) have high f$_x$/f$_{opt}$,  
soft spectra  (characteristic blackbody temperatures kT$\lsim$0.6 keV), 
and low luminosity, similar to the 
AXP (see Fig.~2).  More sensitive searches for pulsations in these sources   (so far   
hampered by the poor statistics) might reveal in the future new AXP (this does 
not apply to the source in Puppis A if the possible periodicity at 75 ms reported 
by  Pavlov et al. (1999) is confirmed by better data).

\section{Models}

Though the absence of a massive companion  
and the presence of a neutron star
are observationally well established,
the AXP remain one of the more enigmatic classes of 
galactic high energy sources. 
Also the main mechanism responsible for the observed X--ray luminosity 
is still  unclear.  
Having excluded models powered by the rotational energy loss
of isolated neutron stars (see section 3.5), the remaining explanations
advanced for the AXP fall into two main classes: models
based on accretion (with or without a binary companion of
very low mass) and those invoking highly magnetized 
neutron stars powered by the decay of the magnetic 
field and/or internal heat dissipation.

Binary models have the advantage of naturally providing
accretion as a  source of energy.
However, the  tight limits on the possible companion 
stars (sections 3.6, 3.7) have also   led to
interpretations based on accretion unto isolated 
neutron stars.


\subsection{Accretion-based Models}

In general, accretion from the interstellar medium  (ISM)
cannot provide the required luminosity under typical  
ISM parameters and neutron star velocities.
In fact, the accretion luminosity is given by
L$_{acc}\sim$10$^{32}$~~v$_{50}^{-3}$~~n$_{100}$~erg s$^{-1}$ 
where  v$_{50}$ is the relative velocity between the
neutron star and the ISM in units of 50 km s$^{-1}$ and
n$_{100}$ is the gas density in units of 100 atoms cm$^{-3}$.
Unless all the AXP lie within nearby ($\sim$100 pc) molecular clouds,
which seems very unlikely considered their distribution
in the galactic plane, the accretion rate   is
clearly insufficient to produce the observed luminosities.
 
van Paradijs et al. (1995) proposed a more efficient scenario,
in which isolated neutron stars are fed from residual accretion disks,
formed after the  complete spiral-in of a neutron star
in the envelope of a giant companion star 
(a Thorne-Zytkow object, TZO, Thorne \& Zytkow 1977).
Thus the AXP could be one possible outcome of
the common envelope evolutionary
phase of close HMXRB systems.
The connection with massive binaries is supported by the
fact that the AXP seem to be relatively young objects,
being located at small distances from the galactic plane
and sometimes found associated with SNR. 
According to van Paradijs et al. (1995), the estimated
birthrate of AXP is consistent with that of TZO.

The idea that AXP are isolated neutron star accreting from a
residual disk has been further developed by Ghosh et al. (1997), who
put this model in the broader context of the evolution 
of close massive binaries. In this scenario, a HMXRB
undergoing common envelope evolution can produce two kinds
of objects, depending on the efficiency with which the 
massive star envelope is lost. 
Relatively wide systems have enough orbital energy to led to
the coplete expulsion of the envelope before the settling of
the neutron star at the center. This would result in the 
formation of binaries composed of a neutron star and a Helium star,
like 4U~1626--67 and Cyg X--3. 
Closer HMXRB, on the other hand, would produce TZO due to 
the complete spiral in of the neutron star in the common envelope
phase. These systems would subsequently evolve into AXP:
isolated neutron stars undergoing accretion from two 
distinct flows: a disk and a spherically symmetric
component, resulting from the part of the envelope with
less angular momentum.
According to Ghosh et al. (1997), this model would also explain
the two component spectra observed in most AXP, as well as    their secular
spin-down:  
the accretion from the   disk is  responsible
for the power-law   and the long term
spin-down due to the decreasing mass accretion rate, while the 
spherically symmetric flow  gives rise to the
blackbody emission from a large fraction of the neutron star surface.

Though this is certainly an interesting model, several 
uncertainties exist. In particular very little is known on
the evolution during the common envelope phase and on the
efficiency of conversion of the orbital binding energy
to that of the dynamical outflow of the envelope.
According to Li (1999), other problems of this model are the
short lifetime of the accretion disk and the fact that
in any case it would be unable to reproduce the 
spin-down behaviour observed in AXP.

Binary models for AXP have not been developed in detail, although
we note that they cannot be completely ruled out in the
case  very low mass companions and/or unfavourable inclination angles
(furthermore, sensitive searches for Doppler modulations have only
been done for three out of six AXP).
In a certain sense, this is the most conservative explanation
since it does not involve new kinds of objects with relatively
uncertain properties. In the context of binary systems
with very low mass companions, Mereghetti \& Stella (1995)
proposed that the AXP are weakly magnetized neutron stars
(B$\sim$10$^{11}$ G) rotating close to the equilibrium  period. 
This requires accretion rates of the order of a few 10$^{15}$ g s$^{-1}$,
consistent with the AXP luminosities.

\subsection{Magnetars}

Models based on 
strongly magnetized (B$\sim$10$^{14}$--10$^{15}$ G) neutron stars, 
or "magnetars", were originally developed to explain the peculiar properties of
SGR (Duncan \& Thompson 1992; Thompson \& Duncan 1995,1996)
and received a substantial support with the discovery
of pulsations and spin-down in these sources (Kouveliotou et al. 1998, 1999,
Hurley et al. 1999).
If one assumes that the AXP spin-down 
is due to  magnetic dipole radiation losses, values of
B = 3.2$\times$10$^{19}$ ($P\pdot$)$^{1/2}$  $\gsim$10$^{14}$ G 
are obtained, suggesting that also 
the X--ray emission from these objects could be powered by magnetic
field decay (see Thompson, these proceedings).

Different authors discussed the kind of  spin-down irregularities
expected in  the magnetar model. Heyl \& Hernquist (1999)
fitted the period histories of \ee and \oo with glitches
similar to those observed in radio pulsars. The same data
were interpreted by Melatos (1999) in terms of a periodic ($\sim$5-10 yrs)
oscillation in $\pdot$ caused by radiative precession,
an effect due to the star asphericity induced by the
very strong magnetic field.
Unfortunately,  the sparse period measurements available
for AXP do not allow for the moment to discriminate
among the different possibilities.

\section{Conclusions}

Though the nature of the AXP is still unknown, after more than
20 years since the discovery of the prototype of this class (\ee),
it is clear that these objects represent an important manifestation 
of neutron stars. There is growing evidence that a large
fraction of neutron stars are born with properties very different
from that of the Crab and Vela pulsars. 
This might explain why only very few energetic, rapidly
spinning radio pulsars have a firm association with a SNR.

Due to their relatively low luminosity and soft spectrum
(critically affected  by the interstellar absorption)
AXP are not easy to find. Several of the       
known X--ray sources, too faint for sensitive pulsations
searches, could be AXP and we can expect that, thanks to 
the coming X--ray satellites, many more will be discovered
in the near future. 
Furthermore, if \axj is confirmed as a ''transient'' AXP,
the overall population of this class of objects would be
even larger than assumed so far.
It might well be that the ''Anomalous'' pulsars are indeed
one of  the 
most ''normal'' manifestations of young neutron stars.


\end{document}